%% file: Carpenter.tex
\documentclass[preprint]{aastex}

\received{}
\accepted{}
\journalid{}{}
\articleid{}{}

\slugcomment{\centerline{\Large\bf \hfil To appear in AJ, May 2001\hfil}}

\shortauthors{Carpenter}
\shorttitle{2MASS Color Transformations}

\newcommand{\ts}{\thinspace}
\newcommand{\simless}{\mathbin{\lower 3pt\hbox
     {$\rlap{\raise 5pt\hbox{$\char'074$}}\mathchar"7218$}}}
\newcommand{\simgreat}{\mathbin{\lower 3pt\hbox
     {$\rlap{\raise 5pt\hbox{$\char'076$}}\mathchar"7218$}}}

\newcommand{\about}    {$\approx$\ts}
\newcommand{\aboutless}{$\simless$\ts}

\newcommand{\JB}{$J$}
\newcommand{\HB}{$H$}
\newcommand{\KB}{$K_s$}
\newcommand{\JH}{$J-H$}
\newcommand{\JK}{$J-K_s$}
\newcommand{\HK}{$H-K_s$}

\newcommand{\M}{$^{\rm m}$}

\newcommand{\etal}{et~al.}

\def\insertplot#1#2#3#4#5#6#7{
\vskip 10pt\nobreak\hbox to \hsize{\hss\dimen0=#3in\hbox to #6\dimen0{%
\dimen0=#2in\vbox to #6\dimen0{\vss
\special{ps: plotfile #1}
\special{ps::[end]
  PGPLOT restore
}
}\hss}\hss}\vskip 10pt}

\begin{document}

\title{Color Transformations for the 2MASS Second Incremental Data Release}

\author{John M. Carpenter\\
        California Institute of Technology, 
        Department of Astronomy, MS 105-24, \\ Pasadena, CA 91125; 
        email: jmc@astro.caltech.edu}

\begin{abstract}

Transformation equations are presented to convert colors and magnitudes 
measured in the AAO, ARNICA, CIT, DENIS, ESO, LCO (Persson standards), MSSSO, 
SAAO, and UKIRT photometric systems to the photometric system inherent to the 
2MASS Second Incremental Data Release. The transformations have been derived 
by comparing 2MASS photometry with published magnitudes and colors for 
stars observed in these systems. Transformation equations have also 
been derived indirectly for the \citet{BB88} and \citet{K83} homogenized 
photometric systems.

\end{abstract}


\section{Introduction}

The 2 Micron All Sky Survey (2MASS) will provide \JB, \HB, and \KB\ band 
photometry for millions of galaxies and nearly a half billion stars. While the
2MASS data alone will make important contributions to many fields of study,
the scientific impact of many programs can be further enhanced 
by comparing 2MASS photometry with existing photometric measurements or by 
conducting follow-up observations. One difficulty in directly comparing 
2MASS photometry with other near-infrared observations is that these comparison
data will often be obtained with a set of filters that have different 
transmissions profiles and effective wavelengths than the filters adopted for 
the 2MASS survey. Variations in the filter transmission characteristics can 
lead to systematic differences in the observed stellar colors, especially for 
objects with extremely red spectral energy distributions or unusual spectral 
line features. Any detailed comparisons between 2MASS data and observations 
conducted at other telescopes then requires that both sets of photometry be 
placed on a common photometric system. Since 2MASS will provide photometry for 
sources over the entire sky, it is natural to adopt the 2MASS photometric 
system as the reference point for these comparisons.

As the 2MASS survey has progressed, subsets of the survey data have been 
publicly released on three occasions to date. The 2MASS Second Incremental 
Release in particular provides photometry for both point and extended sources 
for \about 47\% of the sky. Even though these catalogs are considered 
preliminary, as all data will be reprocessed at the completion of the 
2MASS observations with a final version of the data reduction pipeline, 
the data contained in the Second Incremental Release already meets a high 
degree of global photometric uniformity \citep{Sergei00}. It is therefore
appropriate to determine the color transformations between the 2MASS Second 
Incremental Release and other photometric systems. In this paper, I derive 
these transformation equations using 2MASS observations of the standard stars 
that define commonly used photometric systems. This exercise 
necessarily emphasizes recent grids of faint near-infrared standards 
\citep{CH92,Carter95,Hunt98,Persson98} which have effectively replaced 
more traditional standards \citep{Glass74,Elias82,Allen83,Carter90,BMS91,M94} 
that are usually too bright to be observed with modern instrumentation. The
2MASS data analyzed in this study are summarized in Section~\ref{data}. 
Section~\ref{2mass} examines the internal consistency of the 2MASS photometry,
the temporal stability of the 2MASS photometric system, and the color 
transformations between the northern and southern 2MASS surveys. The color 
transformations between 2MASS and other photometric systems are then presented 
in Section~\ref{transformations}.

\section{2MASS Photometry}
\label{data}

The 2MASS observations are being conducted using 1.3~meter telescopes at 
Mt. Hopkins in Arizona (the Northern survey) and at Cerro Tololo in Chile 
(the Southern survey). Data from both telescopes are processed, calibrated,
and analyzed for quality control at the Infrared Processing and Analysis 
Center (IPAC). The 2MASS 
photometry analyzed in this study has been obtained from the internal 2MASS 
working databases at IPAC that contain both the 2MASS Second Incremental Data 
Release and photometry for the remaining part of the sky observed to date. 
Further, all but one of the 2MASS calibration fields are centered on standard 
stars contained in \citet{Persson98} and \citet{CH92}, and wherever 
possible, the more accurate, globalized 2MASS magnitudes for stars in these 
regions computed by \citet{Sergei00} have been adopted. These additional data 
are included in this analysis to increase the number of standard stars with 
2MASS photometry and allow correspondingly more accurate color transformations 
to be derived. While these additional data meet the photometric requirements 
used to generate the Second Incremental Release catalogs, they may not 
necessarily satisfy other requirements adopted for this release as described 
in the 2MASS Explanatory Supplement \citep{Cutri00}. (For example, the Second
Incremental Release required that all tiles must be part of a block of at
least 3 contiguous tiles in right ascension, a requirement not imposed here.)
Further, it should be noted that the data calibration procedures have evolved 
as the 2MASS survey has progressed, and the derived transformations presented 
in this paper should be regarded as preliminary until the 2MASS data are 
reprocessed in a uniform manner at the completion of survey operations. In 
practice, any changes in the calibration over the course of the survey to date 
have been \aboutless 0.01\M\ (Nikolaev \etal~2000; see also 
Section~\ref{2mass}).

The 2MASS Explanatory Supplement \citep{Cutri00} describes the Second 
Incremental Release in detail, and only relevant aspects of the 
data processing from this document are summarized here. Photometry for 
bright stars (\about 5-9\M) are measured in the 2MASS data reduction pipeline
using a circular aperture on a sequence of six 51~ms integration images. Stars 
brighter than \about 5\M\ are saturated even in these short exposures and 
accurate magnitudes cannot be derived. Magnitudes for fainter stars were 
obtained using Point Spread Function (PSF) fitting photometry on six, 1.3~s 
integration images. The photometric uncertainty for stars measured with 
aperture photometry is computed from the standard deviation of the mean of the 
six aperture magnitudes. For PSF fitting photometry, the photometric 
uncertainty is derived from the Poisson noise associated with the observed 
flux and the fluctuations in the sky background as measured in a sky annulus 
around each star. The normalization to remove any potential systematic 
difference between the PSF and aperture photometry is discussed in the 2MASS 
Explanatory Supplement \citep{Cutri00}. In addition to the internal 
photometric accuracy, a zero-point uncertainty is determined from calibration 
fields observed every hour during survey operations. (For early survey 
operations, the calibration fields were observed every 2 hours.) The total 
photometric uncertainty adopted here is the quadrature sum of the zero-point 
and internal photometric uncertainties for each star. The minimum, total
photometric uncertainties in the \JB, \HB, and \KB\ band magnitudes are 
\about 0.02\M. By comparison, the photometric uncertainties for the 
standard star observations defining the photometric systems discussed in 
Section~\ref{transformations} are \aboutless 0.025\M\ and sometimes less than 
0.01\M. Stars in the overlap region between adjacent 2MASS tiles have more 
than one photometric measurement. While the photometry from only one of these 
observations appears in the public release Point Source Catalog, 
these photometric measurements have been averaged here, with each measurement 
weighted by the inverse square of the total photometric uncertainty. 

\section{The 2MASS Photometric System}
\label{2mass}

\subsection{Filters}
\label{filters}

Both the Northern and Southern 2MASS telescopes are outfitted with a similar 
set of optics, filters, and detectors to observe the \JB, \HB, and \KB\ bands
simultaneously. Since the total transmission through the atmosphere and the
optical elements define any photometric system, it is instructive 
to review these characteristics as they pertain to 2MASS and note any 
substantial differences from other photometric systems. 
Figure~\ref{fig:filters} shows the transmission as a function of wavelength 
through the 2MASS optical path, including the telescope mirror reflectivity, 
the dewar window, anti-reflection coatings, dichroics, filters, and the 
NICMOS detector quantum efficiency, but excluding the atmosphere. The dominant
source of transmission loss through the optical path is the detector quantum
efficiency, which is \about 0.6-0.65 in the \JB, \HB, and \KB\ bandpasses.
A model atmospheric transmission for the mean conditions at Mt. Hopkins is 
shown separately in Figure~\ref{fig:filters} as indicated by the thin solid 
line. The atmospheric transmission data, kindly provided by Martin Cohen, was 
computed using the 
USAF PLEXUS code and binned to a resolution of 0.002\micron\ for display 
purposes. The transmission curves for both the optical elements and the 
atmosphere are tabulated in the 2MASS Explanatory Supplement \citep{Cutri00}. 

The primary distinction between 2MASS and many other photometric systems is 
that the 2MASS \KB\ filter transmission was specially designed to cut off at 
\about 2.3\micron\ in order to reduce the noise contribution from the thermal 
background. In this manner, the noise in the \KB-band observations will be 
less sensitive to variations in the ambient temperature, allowing for a more 
uniform photometric survey. A similar filter has been adopted by the DENIS 
survey \citep{DENIS99} and also was incorporated into the standard star
observations by \citet{Persson98}. By comparison, more traditional Johnson 
$K$ filters have significant transmission out to \about 2.4\micron. Therefore, 
throughout this paper, the 2MASS \KB\ (``K-short'') filter is distinguished
from the Johnson $K$ filter when presenting the color transformations.

As shown in Figure~\ref{fig:filters}, the short and long wavelength cutoff of 
the 2MASS \JB-band filter extends into the atmospheric H$_2$O absorption 
features at \about 1.1\micron\ and 1.4\micron. In dry weather, the 
transmission at these wavelengths can be significant compared 
to typical conditions, implying that the effective \JB-band wavelength, 
calibration zero-points, and possibly color transformations will depend on the 
atmospheric water vapor content. Indeed, the \JB-band calibration zero-points 
often show smooth variations within a night as large as 0.1 magnitudes 
\citep{Cutri00}, and seasonal variations in the average zero-point as large 
as 0.2 magnitudes are observed. The \HB- and \KB-band zero-points are 
relatively constant within a night, although seasonal variations of \about 0.1 
magnitudes are observed. While the 2MASS survey does not record the 
atmospheric water vapor content directly, these zero-point variations are 
presumably due to changes in the amount of water vapor.

The above discussion indicates that various aspects of 2MASS photometry need 
to be investigated before deriving the color transformations between 2MASS
and other photometric systems, namely, 
(1) the temporal stability of the 2MASS photometric system over the 3$+$ years 
    of survey operations; 
(2) the effects of the atmospheric water vapor content on the color 
    transformations, in particular those involving \JB-band;
and 
(3) any differences in the photometric systems between the Northern and
    Southern surveys.
\citet{Sergei00} have discussed these issues in terms of the global magnitude
calibration of the 2MASS survey and found that any temporal changes in the 
global calibration of the \JB, \HB, and \KB\ magnitudes in the 2MASS survey 
are \aboutless 0.01\M. However, since most of the standard stars analyzed by 
\citet{Sergei00} span a small range of colors, the stability of the 2MASS 
photometric system as pertaining to the stellar colors remains to be 
established. The following subsections present such an analysis, and it is 
shown that any internal variations in the 2MASS color transformations are less 
than the photometric uncertainties for any individual star.

\subsection{Temporal Stability}
\label{temporal}

To examine any possible temporal variations in the observed stellar colors, 
2MASS photometry in a \about $1^\circ\times6^\circ$ region near the Galactic 
plane was analyzed that has been observed on two occasions with the Mt. 
Hopkins telescope, once on June 12, 1997 near the start of Northern survey 
operations, and again on May 24, 2000. Thus comparison of these sets of 
observations will indicate any change in the photometry over nearly a 3 year 
time period. Besides the long time baseline, this field was chosen since the 
\JB-band calibration zero-point on the two nights is the same to within 
0.01\M, suggesting that the atmospheric conditions were similar for both sets 
of observations (see Section~\ref{filters}). A similar sized region with large 
differences in the \JB-band zero-points is analyzed in the following section.

Sources were selected from both survey nights that satisfied the following 
criteria:
(1) the source is free of any flags from the 2MASS data processing pipeline
    that indicate the photometry may be contaminated by a nearby star;
and
(2) if the magnitudes were computed using PSF fitting photometry, the reduced
    chi-squared from the PSF fit is $\le$ 2.0 to eliminate any potentially
    extended sources. 
Point sources in the first night of observations that satisfy these criteria 
and contain a signal to noise ratio $\ge$ 15 in each of the \JB, \HB, and \KB\ 
bands were matched with a source in the second observational set using a 
1\arcsec\ search radius. A total of 34,083 sources were matched in this 
manner. The absolute value of the average photometric offset between the two 
nights, computed by averaging the difference in photometry for all matched 
sources, is $0.025$\M, 0.001\M, and $0.002$\M\ at \JB, \HB, and \KB\ band
respectively. Measured relative to the zero-point calibration uncertainties, 
the offsets are 0.8$\sigma$, 0.04$\sigma$, and 0.09$\sigma$. Therefore, any
difference in the photometric zero-points between the two nights is within the
nightly zero-point uncertainties. The zero-point offsets were removed by 
adding a constant to the first night of observations such that the following
analysis compares only the difference of the stellar colors. No offsets have 
been applied to the 2MASS data, however, when comparing the 2MASS photometry 
with other photometric systems. 

Linear fits to the $J-K$, $J-H$, and $H-K$ colors between the two sets of
observations were performed using the FITEXY program \citep{Press92} that 
incorporates uncertainties from both measurements. The reduced chi-squared is 
\about 0.7 for each of the fits, indicating that the colors are well 
modeled by a linear relation and that the residuals are consistent with random 
noise. Based on the fit, the difference in the observed colors between 
observations taken \about 3 years apart is less than 0.01\M\ for $J-H < 1.3$
and $J-K_s < 1.8$. The results for the $H-K_s$ fit are not meaningful since 
the observed dispersion in the colors (0.089\M) is only slightly larger than 
the expected dispersion (0.066\M) due to photometric noise, such that there is 
little intrinsic variations in the $H-K_s$ colors for this field. Nevertheless,
the strong similarity between the $J-H$ and $J-K_s$ colors suggest that any
intrinsic variations in the $H-K_s$ colors must also be small. Thus any 
temporal changes in the 2MASS color system as judged from the $J-H$ and
$J-K_s$ results are comparable to the minimum photometric uncertainties 
(\about 0.02\M) for any individual star.

\subsection{Atmospheric Conditions}
\label{atmosphere}

As discussed in Section~\ref{filters}, the effective transmission through the 
2MASS \JB-band filter depends on the amount of atmospheric water vapor. 
Variations in the water vapor content may lead to changes in the \JB-band 
zero-point calibration and possible introduce additional color terms for the 
survey data. To quantify this effect, 2MASS photometry for another \about 
$1^\circ\times6^\circ$ region was analyzed that has been observed on 2 
occasions by the 
Northern survey telescope in which the \JB-band zero-point between the 2 
nights differed by 0.16\M, presumably due to differences in atmospheric water
vapor content. A total of 49,389 stars were identified between the two 
observations using the criteria described in Section~\ref{temporal}. The 
average photometric offsets between the two observations measured relative to 
the zero-point calibration uncertainties are 0.5$\sigma$, 1.1$\sigma$, and 
1.6$\sigma$ for \JB, \HB, and \KB\ respectively. Thus the photometric offsets 
were consistent within the calibration uncertainties, and the zero-points 
differences were removed before comparing the near-infrared colors (see 
Section~\ref{temporal}). The reduced chi-squared is \about 0.9 for each of the
linear fits to the near-infrared colors, again indicating that the correlation 
between the two sets of observations is well represented by a linear relation.
Based on the fits, the difference in the observed colors between 
the two observations is less than 0.014\M\ for $J-H < 3.2$, less than 0.021\M\ 
for $J-K_s < 4.7$, and less than 0.01\M\ for $H-K_s < 1.5$. Thus no significant 
change relative to the photometric uncertainties for an individual star was 
found in the 2MASS photometric system in varying atmospheric conditions.

\subsection{Comparison between the 2MASS South and 2MASS North Surveys}
\label{north_vs_south}

\begin{figure}
\insertplot{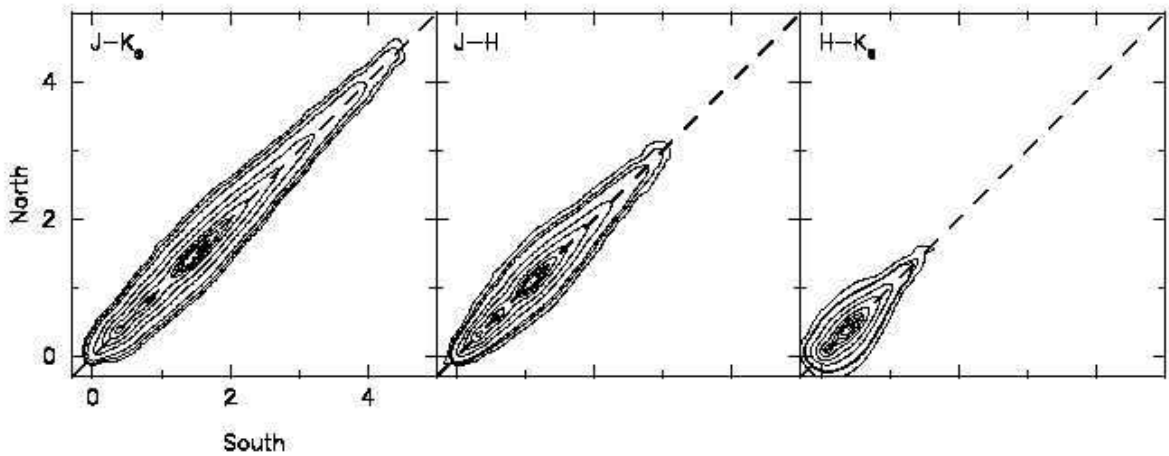}{8.8}{8.1}{0.0}{0.8}{0.75}{1}
\caption{
  Comparison of the \JK, \JH, and \HK\ colors for stars that have 
  been observed by both the 2MASS northern and southern survey telescopes. 
  The density plots were generated by representing each star with a 
  gaussian kernel with a dispersion corresponding to the photometric errors. 
  The contours are at levels of 0.01\%, 0.1\%, 1\%, 5\%, 20\%, and 
  increments of 20\% thereafter of the peak density. The dashed line 
  shows the expected relation if the two colors are equal. 
  \label{fig:2mass}
}
\end{figure}

While much effort has been placed on making the Northern and Southern 
operations as identical as possible, small differences in the color systems 
between the two surveys may still potentially exist.
Any differences in the observed stellar colors between the northern and 
southern hemisphere surveys have been evaluated using 5 fields with 
declinations between $-9$\arcdeg\ and $+3$\arcdeg\ that were observed by both 
telescopes between November 1998 and July 2000. A total of 155,011 point 
sources were matched between the northern and southern survey data using the 
criteria described in Section~\ref{temporal}. The average photometric offset, 
where the offset is in the sense of the magnitude observed in the north minus 
the magnitude observed from the south, ranges from $-0.011$\M\ to 0.039\M\ 
among the three bands. Measured relative to the nightly 1$\sigma$ zero-point 
calibration uncertainties, the offsets range from $-0.45\sigma$ to 
1.9$\sigma$. These photometric zero-point offsets for the 5 fields are within 
the uncertainties of the nightly calibration, and the offsets have been 
applied to the northern data on a night-by-night basis. Figure~\ref{fig:2mass} 
compares the observed \JK, \JH, and \HK\ colors from the northern and southern 
survey telescopes. In each figure, the contours represent the density of points 
in the particular diagram, and the dashed line indicates the expected relation 
if the near-infrared colors are equal. The derived linear relation between the 
two observations are

\begin{eqnarray}
       (J-K_s)_{\rm North} & = & (1.000 \pm 0.001) (J-K_s)_{\rm South} \ \ + \ \ (-0.001 \pm 0.001)\\
         (J-H)_{\rm North} & = & (1.002 \pm 0.001) (J-H)_{\rm South} \ \ + \ \ (-0.003 \pm 0.001)\\
       (H-K_s)_{\rm North} & = & (1.009 \pm 0.001) (H-K_s)_{\rm South} \ \ + \ \ (-0.001 \pm 0.001),
\end{eqnarray}
where the reduced chi-squared from each of the fits is \about 0.9.
These equations indicate that the $J-K_s$ and $J-H$ colors from the Northern
and Southern surveys are statistically indistinguishable. The slope in the 
$H-K_s$ transformation, if significant, implies that the maximum $H-K_s$ color 
difference between the northern and southern survey data over the range of 
colors shown in Figure~\ref{fig:2mass} is 0.012\M, which is within the nightly
calibration uncertainties for any individual star. Therefore, in the remainder 
of this paper, it is assumed that the 2MASS North and South colors systems are 
identical.

\section{Color Transformations}
\label{transformations}

Table~\ref{tbl:standards} summarizes the photometric systems analyzed in the 
paper. Included in the table are the references to the photometry that defines 
these systems and the number of stars with available 2MASS photometry at the 
time of this study that were used to 
derive the color transformations. Tables containing the 2MASS and published
photometry for the individual stars can be found in the 2MASS Explanatory 
Supplement \citep{Cutri00}. Notable omissions from this analysis include the 
homogenized photometric systems put forth by \citet{BB88} and \citet{K83}. The 
\citet{BB88} system is largely based upon the SAAO photometric system 
established by \citet{Glass74}, while \citet{K83} combines the SAAO and 
Johnson \citep{Johnson66} systems. Both the Johnson and Glass standards are 
saturated in the 2MASS images and do not have reliable 2MASS photometry. The 
transformation equations to these homogenized systems have been derived
indirectly as described in the Appendices.

The color transformations between 2MASS and the photometric systems summarized 
in Table~\ref{tbl:standards} were derived by making a linear fit between
the published standard star photometry (or in the case of DENIS, publicly
available catalog data) and the 2MASS observations of these stars. The 
specific variables included in the linear fit are 
$(K_s)_{\rm 2MASS} - K_{std}$ vs. $(J-K)_{std}$, 
$(J-K_s)_{\rm 2MASS}$ vs. $(J-K)_{std}$, 
$(J-H)_{\rm 2MASS}$ vs. $(J-H)_{std}$, 
and
$(H-K_s)_{\rm 2MASS}$ vs. $(H-K)_{std}$, 
where $std$, treated as the X-variable in the fits, represents the photometry 
for the appropriate photometric system. The transformation equations were 
derived using the routine FITEXY \citep{Press92} that minimizes the 
chi-squared between the observations and a straight-line model. The 
uncertainties in both the 2MASS and published photometry are used to evaluate
the chi-squared merit function. After examining the residuals from the fit, 
sources with large discrepancies between the 2MASS and published 
photometry were removed and the fit was re-derived. Any sources removed from 
the analysis are noted below when discussing the results for each photometric 
system. Table~\ref{tbl:q} summarizes the goodness-of-fit parameters from the 
linear fit, including the reduced chi-squared ($\chi^2_\nu$) of the 
residuals and the probability ($q$, $0 \le q \le 1$) that the reduced 
chi-squared can be exceeded by chance for gaussian distributed noise. The 
larger the value of $q$, the more likely the residuals are consistent with 
random noise. Table~\ref{tbl:q} indicates that with the exception of the 
DENIS-2MASS fit (see discussion in Section~\ref{denis}), the 
2MASS and published photometry are reasonably described by a linear 
relationship and the residuals can be explained by photometric noise. The 
residuals for the $K$-band transformations tend to have larger reduced 
chi-squared values than that for the color transformations, especially for 
photometric systems that incorporate very red infrared standards. As discussed 
by \citet[see also Persson \etal~1998]{Elias83}, these red standards tend to 
be near star forming regions and have a greater probability of being variable 
stars. If any of the red standards do have low amplitude variability, the 
magnitude transformations will be most affected since the \JB, \HB, and \KB\ 
band magnitudes will vary simultaneously and produce smaller color changes.

\begin{figure}
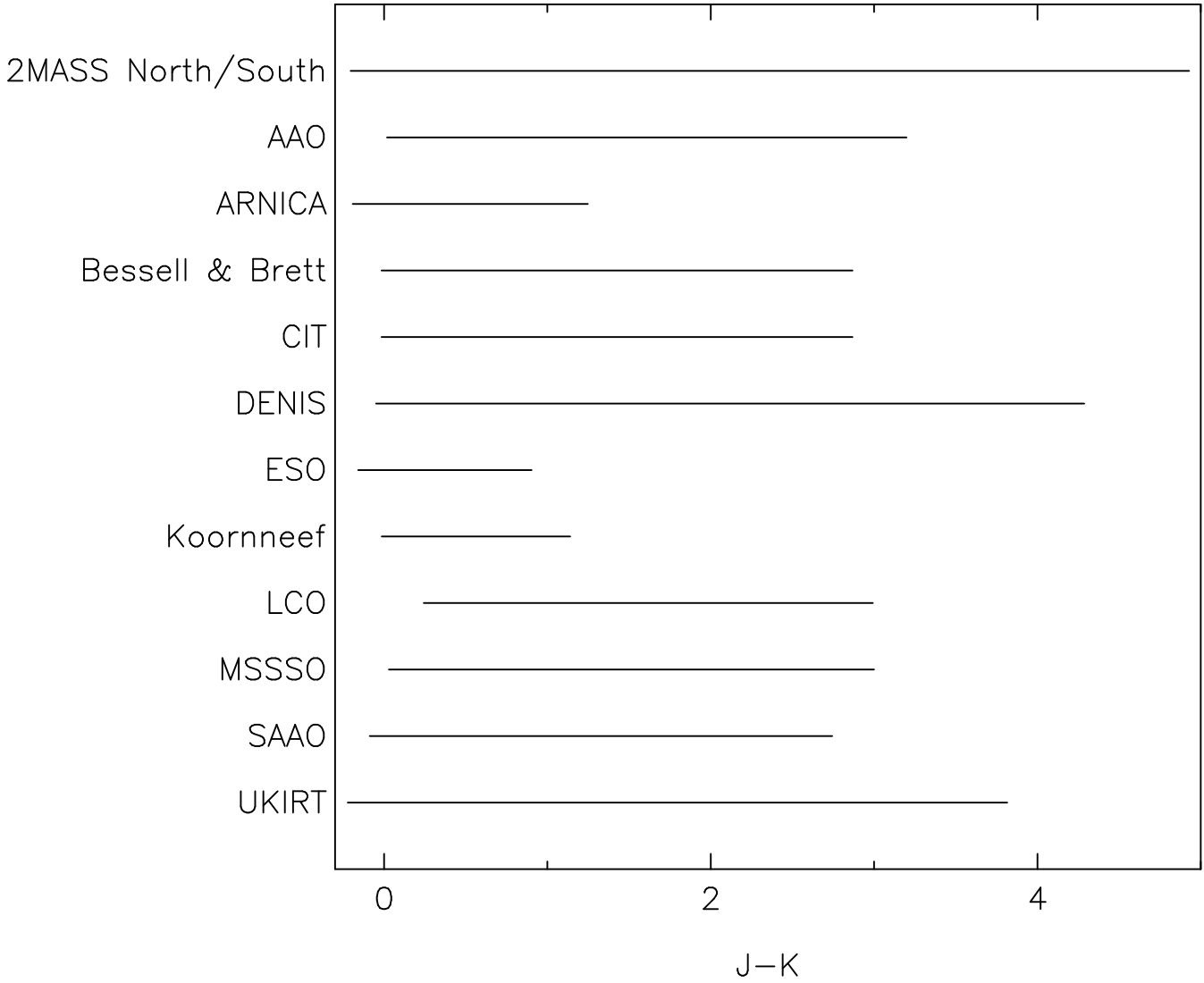

\insertplot{fig03.ps}{6.8}{8.6}{0.0}{2.8}{1.0}{0}
\caption{
  The range of observed $J-K$ colors spanned by the data for the various 
  photometric systems analyzed here. The 2MASS North/South colors refer to the 
  comparison of the 2MASS North and South photometry discussed in 
  Section~\ref{2mass}. The color transformations presented in this paper are 
  not necessarily valid outside the $J-K$ colors shown in this figure. The 
  appropriate limits for the $J-H$ and $H-K$ transformation equations can be 
  obtained from inspection of Figures~\ref{fig:aao}-\ref{fig:saao}.
  \label{fig:jk}
}
\end{figure}

To emphasize that the derived color transformations are valid only for the 
colors spanned by the published photometry, Figure~\ref{fig:jk} shows 
the range of $J-K$ colors contained in the data analyzed here for each 
photometric system. The appropriate ranges for the $J-H$ and $H-K$ colors can 
be obtained from inspection of Figures~\ref{fig:aao}-\ref{fig:koornneef}. In 
addition, these transformations equations may not apply for objects that 
exhibit complex spectral energy distributions (e.g. T dwarfs). For these 
objects, the transformation equations will be sensitive to the exact spectral 
features that are within the filter transmission curve.

The results from the linear fits are summarized graphically in
Figures~\ref{fig:aao}-\ref{fig:koornneef} and are described below for each 
photometric system. In displaying the results, the data are shown as the 
difference between the 2MASS and standard star photometry as a function of the
standard star photometry in order to emphasize subtle, systematic photometric 
differences. This implies that the X and Y axes are correlated in the 
plots, which can create artificial trends with slope of $-1.0$ if the noise in 
the data exceeds the dynamic range in colors. This effect was quite apparent 
in the DENIS-2MASS comparisons since the DENIS data have lower signal to noise 
typically than the 2MASS photometry. For the DENIS results only, the 2MASS 
photometry is plotted along the X-axis.

\subsection{AAO}

\citet{Allen83} present photometric standards for the Anglo-Australian 
Observatory (AAO). Four of these stars are faint enough to be observed with
2MASS and have available 2MASS photometry. In addition, \citet{Elias83} 
obtained photometry for several stars with red near-infrared colors to 
establish the color transformations between the CIT and AAO photometric 
systems. While most of these stars are not considered primary standards, they
were included here since they have a high photometric accuracy (0.01-0.02\M;
Elias \etal~1983). Star cskf-13a in \citet{Elias83} has been observed twice by 
2MASS, and on each occasion the reduced chi-squared from the PSF in each of 
the \JB, \HB, and \KB\ bands is between 2 and 4. Therefore this object is 
extended at the 2MASS resolution and has been removed from the linear fit.

Figure~\ref{fig:aao} shows the correlation between the 2MASS and AAO $K$-band
magnitudes and $J-H$, $H-K$, and $J-K$ colors for the 14 stars with available
photometry. The difference in the $K$-band magnitudes are shown as a function 
of the $J-K$ color in this figure to investigate any color terms in the 
magnitude relations. Two panels are shown for each relation. The top panel 
presents the observed data, and the bottom panel shows the residuals after 
subtracting the fit. The derived color transformations are

\begin{eqnarray}
         (K_s)_{\rm 2MASS} & = & K_{\rm AAO} + (-0.021 \pm 0.008) (J-K)_{\rm AAO} + (-0.032 \pm 0.012)\\
         (J-H)_{\rm 2MASS} & = & (0.924 \pm 0.015) (J-H)_{\rm AAO} \ \ + \ \ (0.000 \pm 0.016)\\
       (J-K_s)_{\rm 2MASS} & = & (0.943 \pm 0.009) (J-K)_{\rm AAO} \ \ + \ \ (0.024 \pm 0.014)\\
       (H-K_s)_{\rm 2MASS} & = & (0.974 \pm 0.033) (H-K)_{\rm AAO} \ \ + \ \ (0.032 \pm 0.016)
\end{eqnarray}

\subsection{ARNICA}

The Arcetri NICMOS3 camera (ARNICA) photometric system is defined by the 
standard observations of 86 northern hemisphere stars by 
Hunt \etal~(1998; see also Hunt \etal~2000).
The calibration of ARNICA system is based on the UKIRT faint standards
\citep{CH92}. The following stars with available 2MASS photometry were omitted 
from the linear fit: 
all five stars in the AS16 group since 3 of the 4 ARNICA standards with 
the highest residuals are in this field;
AS17-4 since higher resolution observations indicate the source is non-stellar
and possibly variable (L. Hunt, private communication);
AS27-0 (UKIRT FS 24) since it is a variable star \citep{H00}; 
AS29-0 (UKIRT FS 25) and AS31-0 (UKIRT FS 28) since they contain a nearby 
companion \citep{H00} that may influence the accuracy of the 2MASS photometry.
Figure~\ref{fig:arnica} compares the 2MASS and ARNICA photometry for 65 stars.
The range of near-infrared colors in the ARNICA standards list is relatively 
small, especially for $H-K$, and further observations are needed with
the ARNICA camera to determine the transformations over a larger range of
near-infrared colors. The derived transformations are

\begin{eqnarray}
         (K_s)_{\rm 2MASS} & = & K_{\rm ARNICA} + (-0.024 \pm 0.011) (J-K)_{\rm ARNICA} + (0.012 \pm 0.006)\\
         (J-H)_{\rm 2MASS} & = & (1.054 \pm 0.020) (J-H)_{\rm ARNICA} \ \ + \ \ (-0.025 \pm 0.008)\\
       (J-K_s)_{\rm 2MASS} & = & (1.056 \pm 0.016) (J-K)_{\rm ARNICA} \ \ + \ \ (-0.018 \pm 0.008)\\
       (H-K_s)_{\rm 2MASS} & = & (1.059 \pm 0.070) (H-K)_{\rm ARNICA} \ \ + \ \ (0.007 \pm 0.008)
\end{eqnarray}

\subsection{CIT}
\label{cit}

The Caltech (CIT) photometric system was described by
\citet{Frogel78} and became defined by the standard star observations from
\citet{Elias82}. In addition, \citet{Elias83} obtained photometry in the CIT
system for a number of stars with large near-infrared colors. As noted above, 
cskf-13a from \citet{Elias83} has been omitted from this analysis. 
Figure~\ref{fig:cit} compares the photometry for 41 stars with CIT and 2MASS
measurements. The derived color transformations are

\begin{eqnarray}
         (K_s)_{\rm 2MASS} & = & K_{\rm CIT} + (0.000 \pm 0.005) (J-K)_{\rm CIT} + (-0.024 \pm 0.003)\\
         (J-H)_{\rm 2MASS} & = & (1.076 \pm 0.010) (J-H)_{\rm CIT} \ \ + \ \ (-0.043 \pm 0.006)\\
       (J-K_s)_{\rm 2MASS} & = & (1.056 \pm 0.006) (J-K)_{\rm CIT} \ \ + \ \ (-0.013 \pm 0.005)\\
       (H-K_s)_{\rm 2MASS} & = & (1.026 \pm 0.020) (H-K)_{\rm CIT} \ \ + \ \ (0.028 \pm 0.005)
\end{eqnarray}

\subsection{DENIS}
\label{denis}

The DENIS survey has released preliminary $I$, \JB, and \KB\ band photometry 
for \about 2\% of the southern sky \citep{DENIS99}. Data from the DENIS survey 
was cross correlated with the 2MASS data for regions near the Chamaeleon I 
dark cloud and near the Galactic Plane at a longitude of \about 323\arcdeg\ in 
order to find select stars that have large extinctions and correspondingly 
large near-infrared colors. No attempt was made to assess the spatial 
uniformity of the DENIS data and the color transformations across the sky. 
Using the data server at the Centre de Donn\'ees astronomiques de Strasbourg, 
DENIS point sources were selected that had at least a signal to noise ratio of 
15 in both the \JB\ and \KB\ bands with no error flags. DENIS sources were 
matched with 2MASS counterparts with a 5\arcsec\ search radius. Ten stars had 
2MASS and DENIS \KB-band photometry that differed by more than 0.5\M\ and 
were omitted from the fit. The 2MASS and DENIS photometry for the 190 remaining
stars is shown in Figure~\ref{fig:denis}. The 2MASS and DENIS photometry are
evidently correlated, but as indicated in Table~\ref{tbl:q}, the reduced 
chi-squared from the linear fit is significantly greater than expected based 
on random noise. This may be a result from comparing photometry in crowded 
regions in the Galactic plane, which was necessary in order to identify red 
stars for the color transformations. The large reduced chi-squared values 
indicate that the uncertainties in the fitted parameters quoted below are
underestimated, perhaps by as much as a \about $\sqrt{2}$ given the value
of $\chi^2_\nu$. The derived transformations are 
\begin{eqnarray}
         (K_s)_{\rm 2MASS} & = & (K_s)_{\rm DENIS} + (0.006 \pm 0.004) (J-K_s)_{\rm DENIS} + (-0.024 \pm 0.006)\\
       (J-K_s)_{\rm 2MASS} & = & (0.981 \pm 0.006) (J-K_s)_{\rm DENIS} \ \ + \ \ (0.023 \pm 0.009)\\
\end{eqnarray}

\subsection{ESO}

The European Southern Observatory (ESO) photometric system was defined
by \citet{Engels81} and \citet{W81} and later updated by \citet{BMS91}
and \citet{BMB96}. Figure~\ref{fig:eso} compares the ESO and 2MASS 
photometry for 56 standards in \citet{BMB96} that have available 2MASS 
photometry. As seen in this figure, the range of near-infrared colors spanned 
by the available data is limited. The derived color transformations are

\begin{eqnarray}
         (K_s)_{\rm 2MASS} & = & K_{\rm ESO} + (0.005 \pm 0.011) (J-K)_{\rm ESO} + (-0.045 \pm 0.004)\\
         (J-H)_{\rm 2MASS} & = & (0.983 \pm 0.030) (J-H)_{\rm ESO} \ \ + \ \ (-0.049 \pm 0.008)\\
       (J-K_s)_{\rm 2MASS} & = & (0.956 \pm 0.017) (J-K)_{\rm ESO} \ \ + \ \ (-0.008 \pm 0.006)\\
       (H-K_s)_{\rm 2MASS} & = & (0.956 \pm 0.126) (H-K)_{\rm ESO} \ \ + \ \ (0.034 \pm 0.006)
\end{eqnarray}

\subsection{LCO (Persson standards)}

\citet{Persson98} developed a grid of $J$, $H$, $K$, and \KB-band standards
for the HST NICMOS camera using observations from the Las Campanas Observatory
(LCO) in Chile. Included in the Persson list of standards are many of the red
stars observed by \citet{Elias83} in the CIT and AAO photometric systems. As 
discussed above, star cskf-13a in the \citet{Elias83} list have been omitted.
In addition, the 2MASS photometry for IRAS 537s differs from the 
\citet{Persson98} photometry by up to 0.26\M, and L547 is a variable star 
\citep{Persson98}. Excluding these three stars, 82 stars from \citet{Persson98} 
have available 2MASS photometry at the time of this study. It should be noted 
that a \citet{Persson98} standard has been adopted as the fiducial calibrator 
in 29 of the 35 2MASS calibration fields, thereby connecting the 
zero-points between the 2MASS and LCO photometric systems. A photometric 
offset of roughly zero, then, is expected in the 2MASS-LCO transformation 
equations, although not necessarily a unit slope. Indeed, as shown in 
Figure~\ref{fig:lco} for the LCO $K$-band data and Figure~\ref{fig:lco_s} for 
the LCO \KB-band data, the zero-point offset between the LCO and 2MASS 
photometry is approximately zero. The derived transformations for the LCO $K$ 
and \KB\ band photometry are

\begin{eqnarray}
         (K_s)_{\rm 2MASS} & = & K_{\rm LCO} + (-0.001 \pm 0.002) (J-K)_{\rm LCO} + (-0.006 \pm 0.004)\\
         (J-H)_{\rm 2MASS} & = & (0.995 \pm 0.006) (J-H)_{\rm LCO} \ \ + \ \ (0.002 \pm 0.006)\\
       (J-K_s)_{\rm 2MASS} & = & (1.013 \pm 0.005) (J-K)_{\rm LCO} \ \ + \ \ (-0.007 \pm 0.006)\\
       (H-K_s)_{\rm 2MASS} & = & (1.008 \pm 0.010) (H-K)_{\rm LCO} \ \ + \ \ (0.002 \pm 0.005)
\\\nonumber
\\
         (K_s)_{\rm 2MASS} & = & (K_s)_{\rm LCO} + (-0.002 \pm 0.002) (J-K_s)_{\rm LCO} + (-0.010 \pm 0.004)\\
       (J-K_s)_{\rm 2MASS} & = & (1.007 \pm 0.005) (J-K_s)_{\rm LCO} \ \ + \ \ (0.002 \pm 0.006)\\
       (H-K_s)_{\rm 2MASS} & = & (1.019 \pm 0.010) (H-K_s)_{\rm LCO} \ \ + \ \ (0.005 \pm 0.005)
\end{eqnarray}

\subsection{MSSSO}

\citet{M94} presented standard star photometry for the Mount Stromlo 
and Siding Spring Observatory (MSSSO), which is nearly identical to the Mount 
Stromlo Observatory (MSO) photometric system described by \citet{JH82}.
In addition, \citet{M94} report photometry in the MSSSO system for a number
of red stars. These stars were included when deriving the color 
transformations in order to increase the accuracy of the transformations. Star
cskf-13a was excluded from this analysis as already noted, as well as the 
stars from \citet{JH80} as they are located in crowded regions in the Galactic 
Plane. Figure~\ref{fig:mssso} compares the MSSSO and 2MASS photometry for the 
20 stars from \citet{M94} used in the analysis. The derived color 
transformations are

\begin{eqnarray}
         (K_s)_{\rm 2MASS} & = & K_{\rm MSSSO} + (-0.021 \pm 0.006) (J-K)_{\rm MSSSO} + (-0.023 \pm 0.008)\\
         (J-H)_{\rm 2MASS} & = & (0.991 \pm 0.014) (J-H)_{\rm MSSSO} \ \ + \ \ (-0.010 \pm 0.014)\\
       (J-K_s)_{\rm 2MASS} & = & (1.005 \pm 0.008) (J-K)_{\rm MSSSO} \ \ + \ \ (0.011 \pm 0.011)\\
       (H-K_s)_{\rm 2MASS} & = & (1.037 \pm 0.029) (H-K)_{\rm MSSSO} \ \ + \ \ (0.019 \pm 0.012)
\end{eqnarray}

\subsection{SAAO}
\label{saao}

\citet{Carter90} updated the South Africa Astronomical Observatory (SAAO) 
photometric system originally defined by \citet{Glass74} and later 
extended the standard star list to fainter magnitudes \citep{Carter95}. 
Figure~\ref{fig:saao} compares the SAAO and 2MASS photometry for 94 stars. 
Most of the relatively blue stars are from the list of $K$\about 8 standards 
in \citet{Carter95}, while the stars with red colors are predominantly from 
\citet{Carter90}. The derived transformations are

\begin{eqnarray}
         (K_s)_{\rm 2MASS} & = & K_{\rm SAAO} + (0.020 \pm 0.007) (J-K)_{\rm SAAO} + (-0.025 \pm 0.004)\\
         (J-H)_{\rm 2MASS} & = & (0.949 \pm 0.018) (J-H)_{\rm SAAO} \ \ + \ \ (-0.054 \pm 0.006)\\
       (J-K_s)_{\rm 2MASS} & = & (0.940 \pm 0.010) (J-K)_{\rm SAAO} \ \ + \ \ (-0.011 \pm 0.005)\\
       (H-K_s)_{\rm 2MASS} & = & (0.961 \pm 0.036) (H-K)_{\rm SAAO} \ \ + \ \ (0.040 \pm 0.005)
\end{eqnarray}

\subsection{UKIRT}

The standard star photometry for the United Kingdom Infrared Telescope (UKIRT) 
was first established by \citet{CH92} and extended by \citet{H00}. Three 
of the 2MASS calibration fields include five UKIRT standards (FS4, FS13, and 
FS15-17; Nikolaev \etal~2000). The UKIRT photometric system is being replaced 
by the Mauna Kea Observatory near-infrared photometric system, which is a 
result of a coordinated effort by several observatories to use a consistent 
set of filters. Fundamental photometry of the standard stars in this new 
photometric system was not available at the time of this study. Therefore, 
the color transformations between the UKIRT and 2MASS photometric systems are 
derived here, and the most recent transformations between the UKIRT and 
Mauna Kea Observatory systems can be found in \citet{H00}. 

Star FS18 was omitted from the linear fit since it is a unresolved
double system at the resolution of 2MASS with 1.38\arcsec\ resolution 
\citep{H00}. Also, the 2MASS \JB-band photometry for FS142 and FS143 differ 
from the UKIRT magnitudes by +0.30 magnitudes and $-0.42$ magnitudes 
respectively, although the \HB\ and \KB\ magnitudes are not significantly 
discrepant. Both stars are located in the Serpens star forming region and have 
$(J-K_s)$ colors between 2.5 and 3.5 magnitudes, and one (FS142, also known as 
EC51) is a possible variable star \citep{Kaas99}. It is unclear then if the 
\JB-band magnitudes for these two stars is a result of low signal to noise 
($<$ 10) in the 2MASS data, or if these are variable stars. Both objects were 
removed from the fit. Figure~\ref{fig:ukirt} compares the UKIRT 
and 2MASS photometry for 72 stars. The derived transformations are

\begin{eqnarray}
         (K_s)_{\rm 2MASS} & = & K_{\rm UKIRT} + (0.004 \pm 0.006) (J-K)_{\rm UKIRT} + (0.002 \pm 0.004)\\
         (J-H)_{\rm 2MASS} & = & (1.069 \pm 0.015) (J-H)_{\rm UKIRT} \ \ + \ \ (-0.027 \pm 0.007)\\
       (J-K_s)_{\rm 2MASS} & = & (1.069 \pm 0.011) (J-K)_{\rm UKIRT} \ \ + \ \ (-0.012 \pm 0.006)\\
       (H-K_s)_{\rm 2MASS} & = & (1.062 \pm 0.027) (H-K)_{\rm UKIRT} \ \ + \ \ (0.017 \pm 0.005)
\end{eqnarray}

\section{Summary}
\label{summary}

Colors and magnitudes of stars observed in the AAO, ARNICA, CIT, DENIS, ESO,
LCO, MSSSO, SAAO, and UKIRT photometric systems are compared to 
photometry from the 2MASS survey. These data have been used to derive the 
$K$-band, $J-H$, $J-K$, and $H-K$ color transformations from these systems to 
the photometric system implicit to the 2MASS Second Incremental Data Release. 
The transformation equations for the \citet{BB88} and \citet{K83} homogenized 
photometric systems have been derived indirectly by first transforming their 
results to the CIT and SAAO systems respectively, and then to the 2MASS system.
The range of colors that these transformation equations are valid over is set 
by the published photometry and can be determined through inspection of 
Figures~\ref{fig:jk}-\ref{fig:koornneef}. 

\acknowledgements

JMC would like to thank Mike Skrutskie for many useful suggestions throughout
the course of this work. He also thanks Roc Cutri, Sandy Leggett, Jay Elias,
and the anonymous referee for their comments on this paper, and Bill Wheaton 
and Martin Cohen for their assistance in obtaining the data for the filter 
transmissions and atmospheric models. This publication makes use of data 
products from the Two Micron All 
Sky Survey, which is a joint project of the University of Massachusetts and 
the Infrared Processing and Analysis Center, funded by the National 
Aeronautics and Space Administration and the National Science Foundation. 
2MASS science data and information services were provided by the InfraRed 
Science Archive (IRSA) at IPAC. This research has made use of the SIMBAD 
database, operated at CDS, Strasbourg, France. JMC acknowledges support from 
Long Term Space Astrophysics Grant NAG5-8217 and the Owens Valley Radio 
Observatory, which is supported by the National Science Foundation through NSF 
grant number AST-9981546.

\appendix

\section{Bessell \& Brett Homogenized System}

\citet{BB88} present a homogenized photometric system based largely on
the SAAO observations from \citet{Glass74}. The \citet{Glass74} standard
stars are saturated in the 2MASS images and cannot be tied directly to the
2MASS photometric system. However, the \citet{BB88} and 2MASS photometric
systems can be related indirectly using the transformations derived here
and those listed in \citet{BB88}. In particular, the CIT transformation 
equations were used since they have been derived using stars that span a 
large range of near-infrared colors. Formally, the transformation between the 
$K$ magnitude measured in the CIT system and the \citet{BB88} homogenized 
system contains a weak dependence on the $V-K$ stellar color. This term was
not included in deriving the following transformations:

\begin{eqnarray}
    (K_s)_{\rm 2MASS}   & = & K_{\rm BB} + (0.000 \pm 0.005)(J-K)_{\rm BB} + (-0.044 \pm 0.003)\\
    (J-H)_{\rm 2MASS}   & = & (0.980 \pm 0.009) (J-H)_{\rm BB} \ \ + \ \ (-0.045 \pm 0.006)\\
    (J-K_s)_{\rm 2MASS} & = & (0.972 \pm 0.006) (J-K)_{\rm BB} \ \ + \ \ (-0.011 \pm 0.005)\\
    (H-K_s)_{\rm 2MASS} & = & (0.996 \pm 0.019) (H-K)_{\rm BB} \ \ + \ \ (0.028 \pm 0.005)
\end{eqnarray}

\section{Koornneef Homogenized System}

\begin{figure}
\insertplot{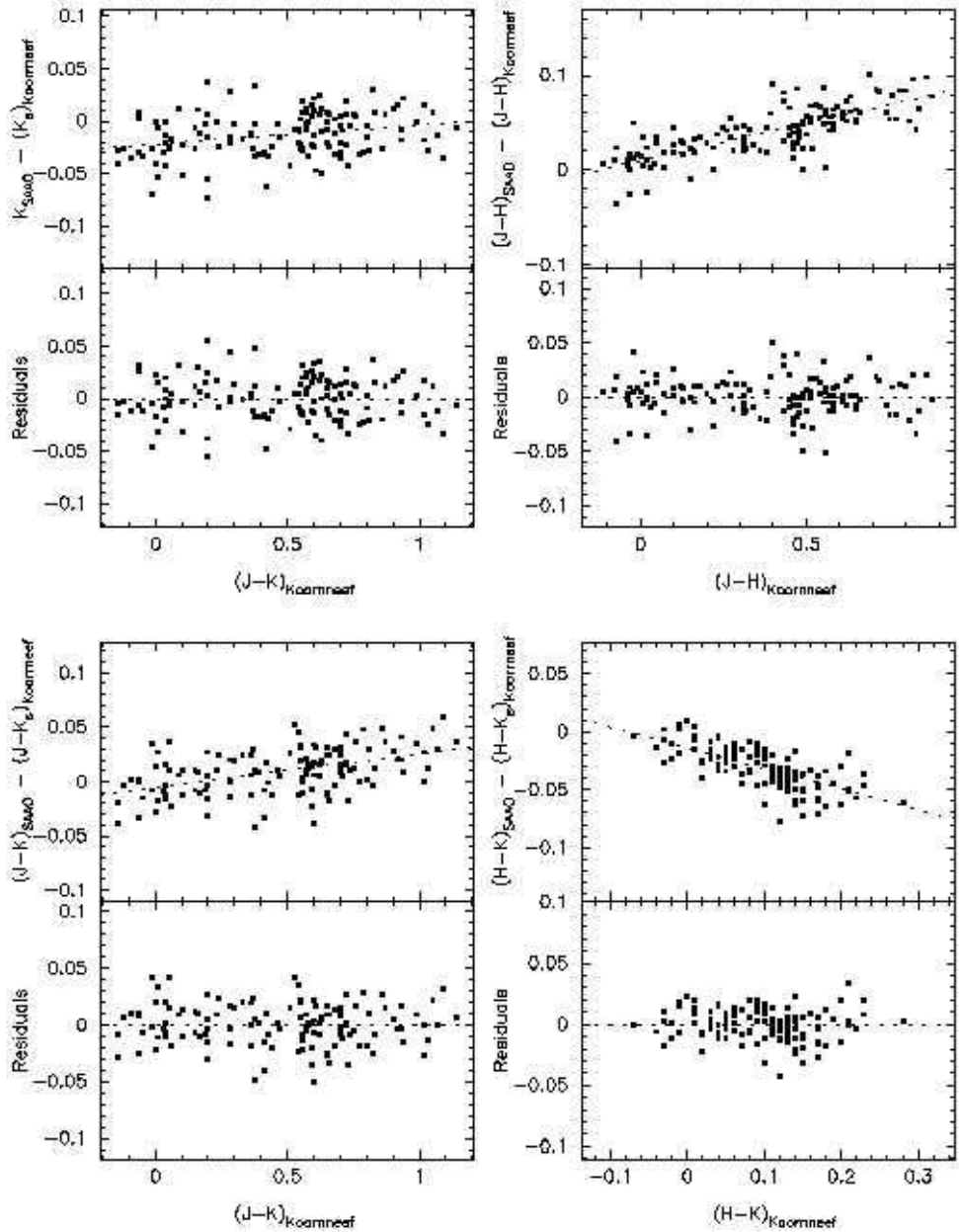}{8.8}{8.1}{0.0}{0.8}{0.8}{0}
\caption{
  Comparison of the photometry for 133 stars in common between 
  \citet{Carter90} and \citet{K83}. The error bars for the individual
  points have been omitted for clarity. These results were used in
  combination with the 2MASS-SAAO color transformations derived
  in Section~\ref{saao} to determine indirectly the transformations
  between the \citet{K83} homogenized photometric system and 2MASS.
  \label{fig:koornneef}
}
\end{figure}

\citet{K83} present an homogenized photometric system that is a hybrid of the 
Johnson \citep{Johnson66} and SAAO \citep{Glass74} systems. Nearly all of
the standard stars listed in \citet{K83} are saturated in the 2MASS data,
and transformations to the 2MASS photometric system could not be derived
directly. However, a number of stars in \citet{K83} were also observed by
\citet{Carter90} in the SAAO system, and can be used to tie the \citet{K83} 
system to 2MASS using the results from Section~\ref{saao}. 
Figure~\ref{fig:koornneef} compares the \citet{Carter90} and \citet{K83} 
photometry for 133 stars in common between the two studies. An uncertainty of 
0.02\M\ was assumed for the \citet{K83} photometry in performing the linear
fit using the FITEXY routine \citep{Press92}. The derived color 
transformations between the two systems are
\begin{eqnarray}
         K_{\rm SAAO} & = & K_{\rm Koornneef} + (0.018 \pm 0.008) (J-K)_{\rm Koornneef} + (-0.022 \pm 0.004)\\
         (J-H)_{\rm SAAO} & = & (1.079 \pm 0.014) (J-H)_{\rm Koornneef} \ \ + \ \ (0.010 \pm 0.006)\\
       (J-K)_{\rm SAAO} & = & (1.032 \pm 0.011) (J-K)_{\rm Koornneef} \ \ + \ \ (-0.006 \pm 0.006)\\
       (H-K)_{\rm SAAO} & = & (0.824 \pm 0.050) (H-K)_{\rm Koornneef} \ \ + \ \ (-0.014 \pm 0.006)
\end{eqnarray}
Using the 2MASS-SAAO results from Section~\ref{saao}, the color transformations
between the \citet{K83} and 2MASS systems are
\begin{eqnarray}
     (K_s)_{\rm 2MASS}   & = & K_{\rm Koornneef} + (0.039 \pm 0.019) (J-K)_{\rm Koornneef} + (-0.047 \pm 0.006)\\
     (J-H)_{\rm 2MASS}   & = & (1.024 \pm 0.024) (J-H)_{\rm Koornneef} \ \ + \ \ (-0.045 \pm 0.006)\\
     (J-K_s)_{\rm 2MASS} & = & (0.970 \pm 0.015) (J-K)_{\rm Koornneef} \ \ + \ \ (-0.017 \pm 0.005)\\
     (H-K_s)_{\rm 2MASS} & = & (0.792 \pm 0.056) (H-K)_{\rm Koornneef} \ \ + \ \ (0.027 \pm 0.005)
\end{eqnarray}

\clearpage

\input{table1}
\input{table2}

\begin{figure}
\insertplot{fig01.ps}{6.8}{8.6}{0.0}{1.8}{1.0}{0}
\caption{
  Transmission curves for the 2MASS optical path (thick solid curves), 
  including the telescope mirror reflectivity, dewar window, anti-reflection 
  coatings, dichroics, filters, and the NICMOS detector 
  quantum efficiency, but excluding atmospheric absorption. The thin solid 
  line shows the model atmospheric transmission for the mean observing 
  conditions at Mt. Hopkins computed using the USAF PLEXUS code and binned
  to a resolution of 0.002\micron.
  \label{fig:filters}
}
\end{figure}
\clearpage

\clearpage

\clearpage

\begin{figure}
\insertplot{fig04.ps}{8.8}{8.1}{0.0}{0.8}{0.8}{0}
\caption{
  Comparison of the photometry of stars observed by \citet{Allen83} and 
  \citet{Elias83} in the AAO photometric system that have available 2MASS 
  photometry. The vertical and horizontal bars indicate the 1$\sigma$
  photometric uncertainties, although the AAO uncertainties are often smaller
  then the symbol size given the dynamic range along the X-axis. The upper 
  left panel plots the difference in the $K$-band magnitudes as a function of 
  the AAO $J-K$ color, and the remaining panels directly compare the $J-H$, 
  $J-K$, and $H-K$ colors. The dotted line in the upper portion of each panel 
  shows the derived transformation between the AAO and 2MASS photometric 
  systems. The bottom portion of each panel shows the residuals from the fit, 
  where the horizontal dotted line at zero is drawn for reference.
  \label{fig:aao}
}
\end{figure}
\clearpage

\begin{figure}
\insertplot{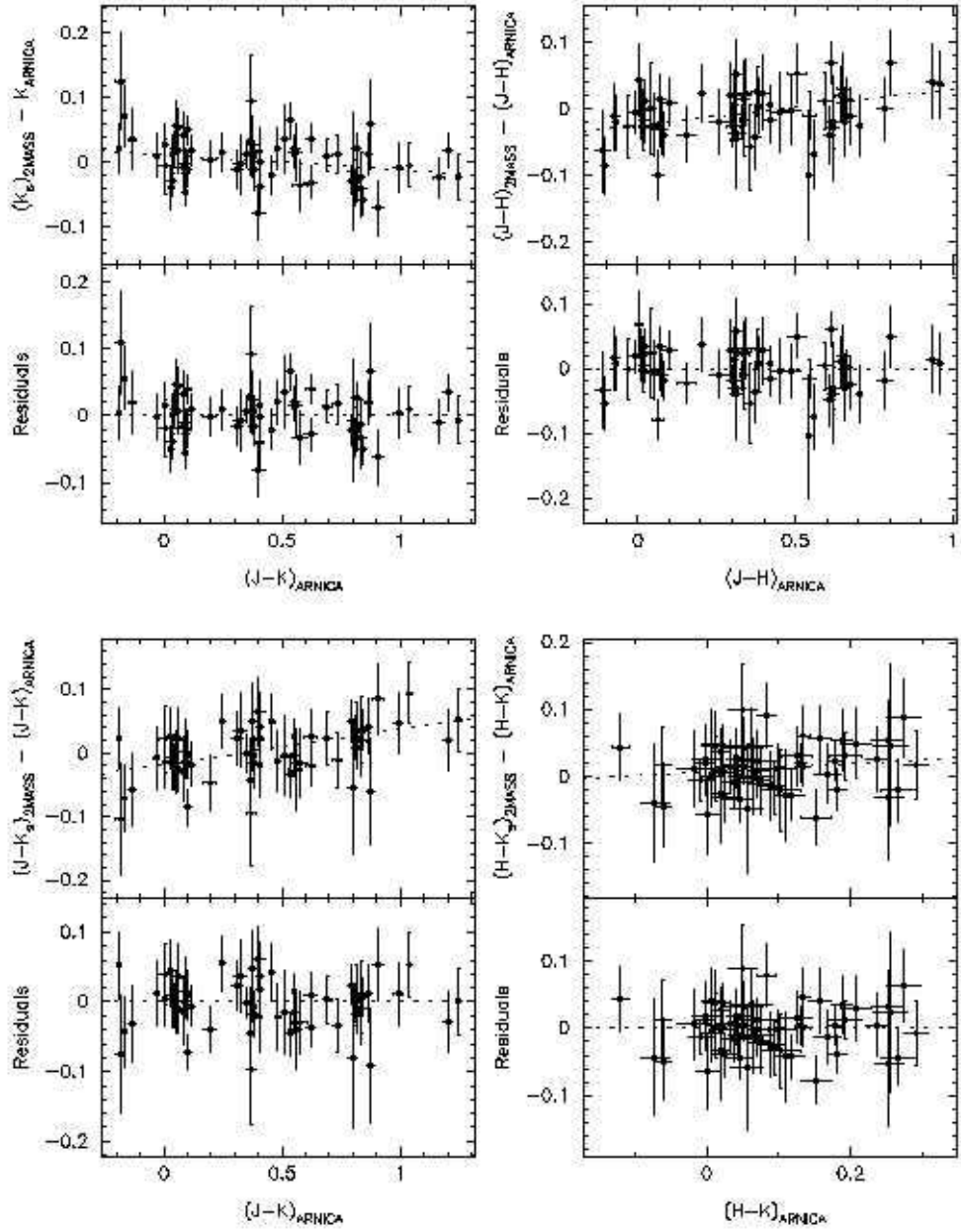}{8.8}{8.1}{0.0}{0.8}{0.8}{0}
\caption{
  Same as Figure~\ref{fig:aao}, except for the ARNICA photometric system. The
  ARNICA standard star photometry is from \citet{Hunt98}. 
  \label{fig:arnica}
}
\end{figure}
\clearpage

\begin{figure}
\insertplot{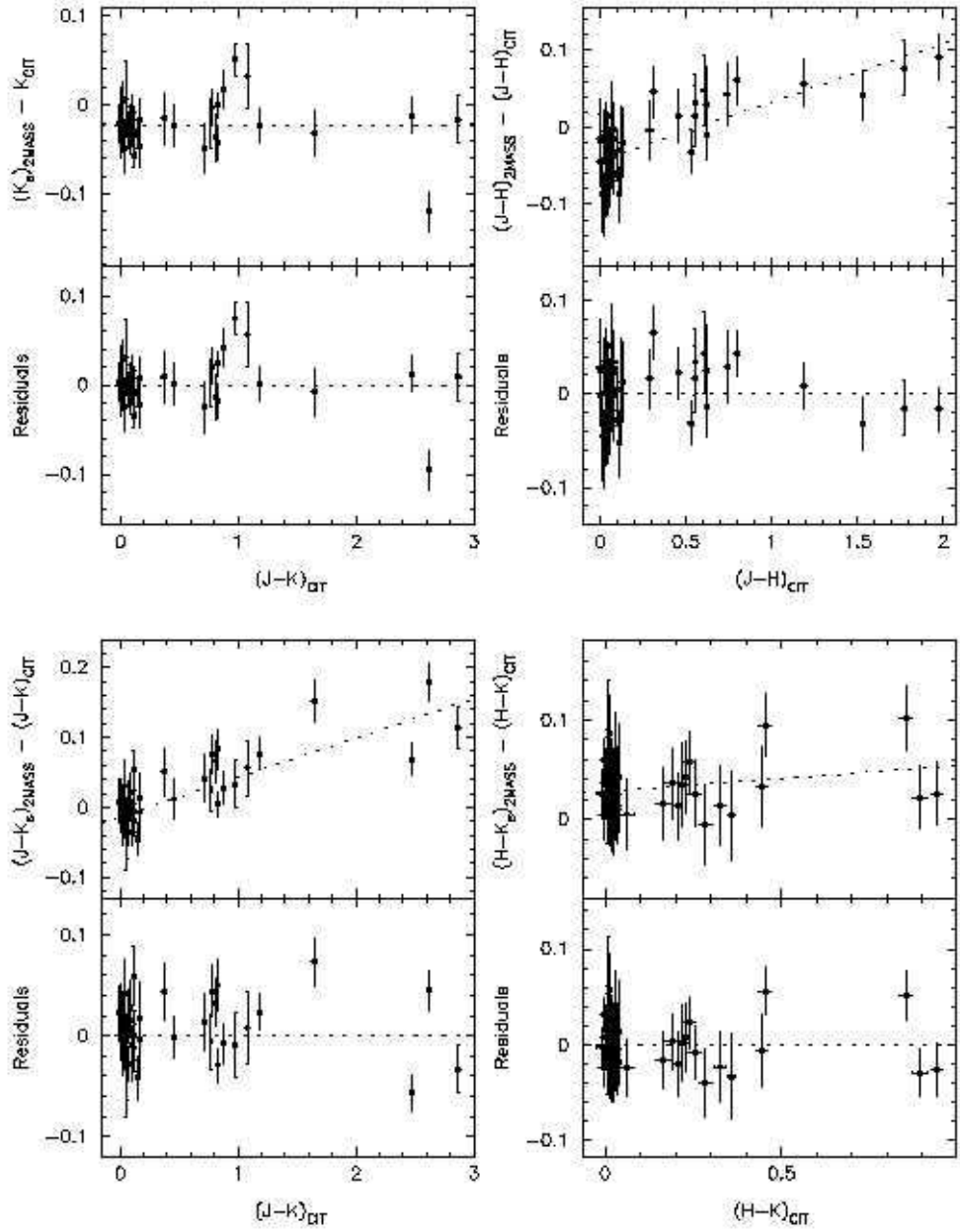}{8.8}{8.1}{0.0}{0.8}{0.8}{0}
\caption{
  Same as Figure~\ref{fig:aao}, except for the CIT photometric system. The
  CIT photometry is from \citet{Elias82} and \citet{Elias83}.
  \label{fig:cit}
}
\end{figure}
\clearpage

\begin{figure}
\insertplot{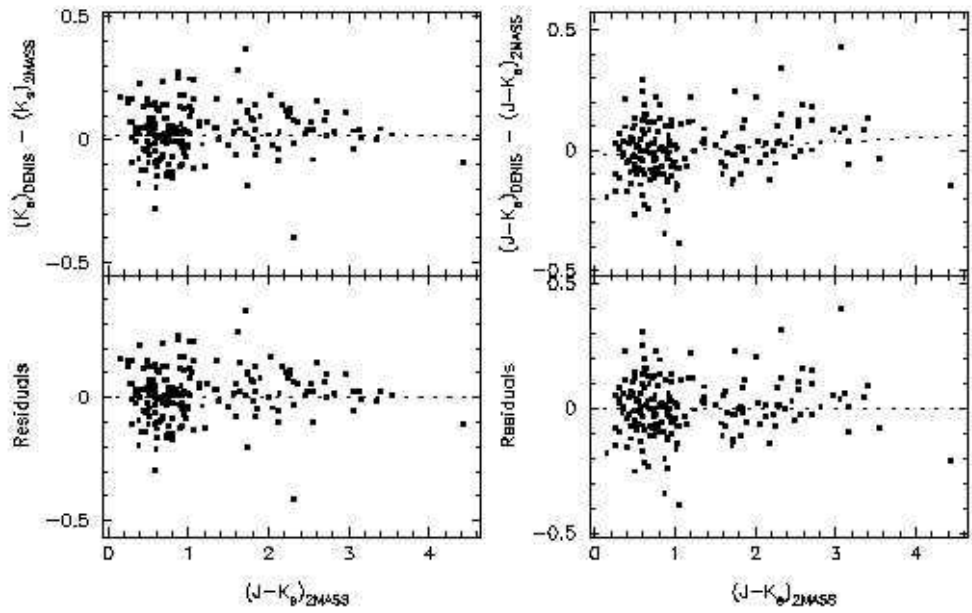}{8.8}{8.1}{0.0}{0.8}{0.8}{0}
\caption{
  Same as Figure~\ref{fig:aao}, except for the DENIS photometric system. 
  Note that unlike other comparisons to published photometry presented here,
  the correlations are plotted as a function of the 2MASS photometry. The 
  error bars for the individual points have been omitted for clarity. The 
  DENIS photometry is from the DENIS preliminary data release described by 
  \citet{DENIS99}.
  \label{fig:denis}
}
\end{figure}
\clearpage

\begin{figure}
\insertplot{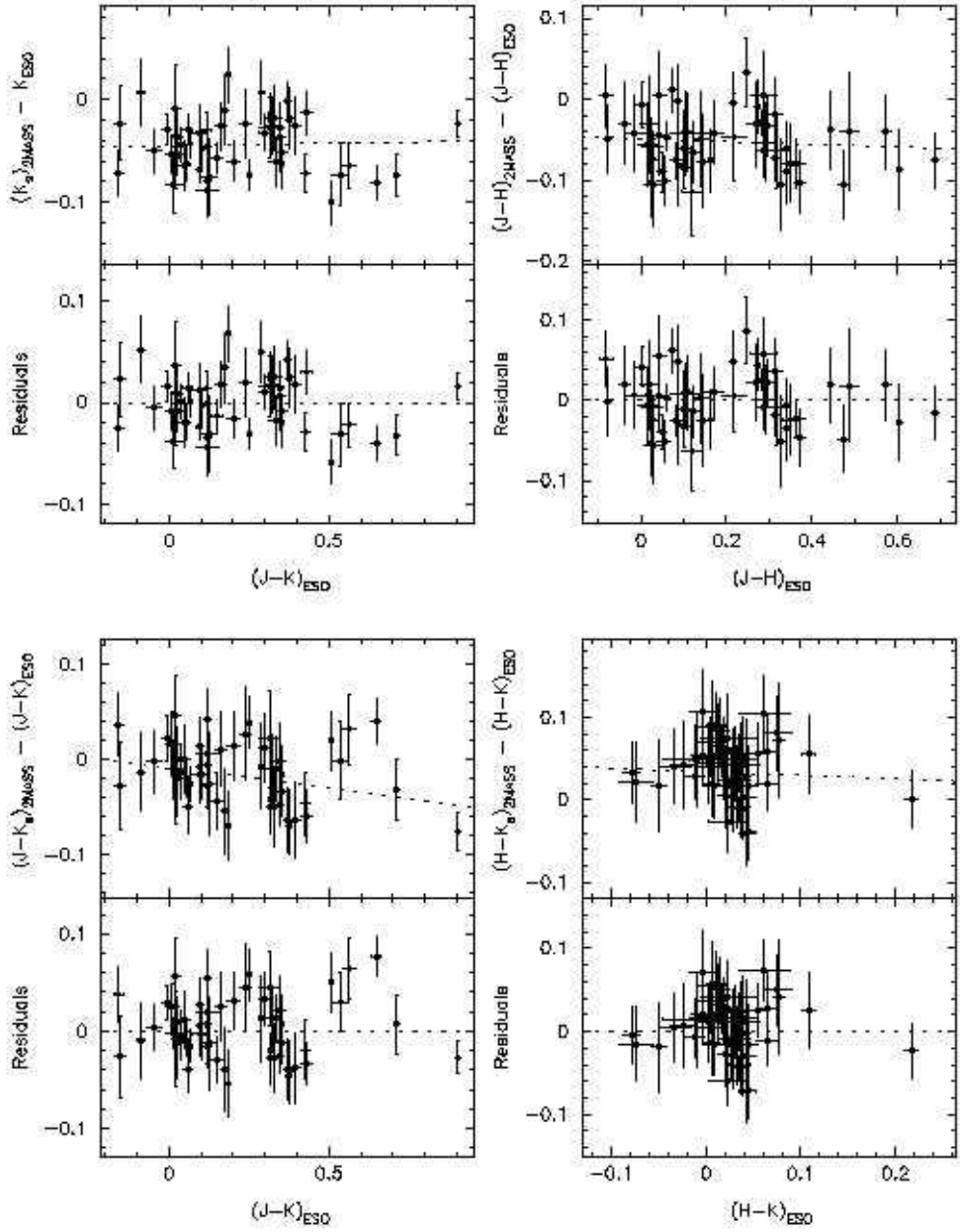}{8.8}{8.1}{0.0}{0.8}{0.8}{0}
\caption{
  Same as Figure~\ref{fig:aao}, except for the ESO photometric system. The
  ESO standard star photometry is from \citet{BMB96}.
  \label{fig:eso}
}
\end{figure}
\clearpage

\begin{figure}
\insertplot{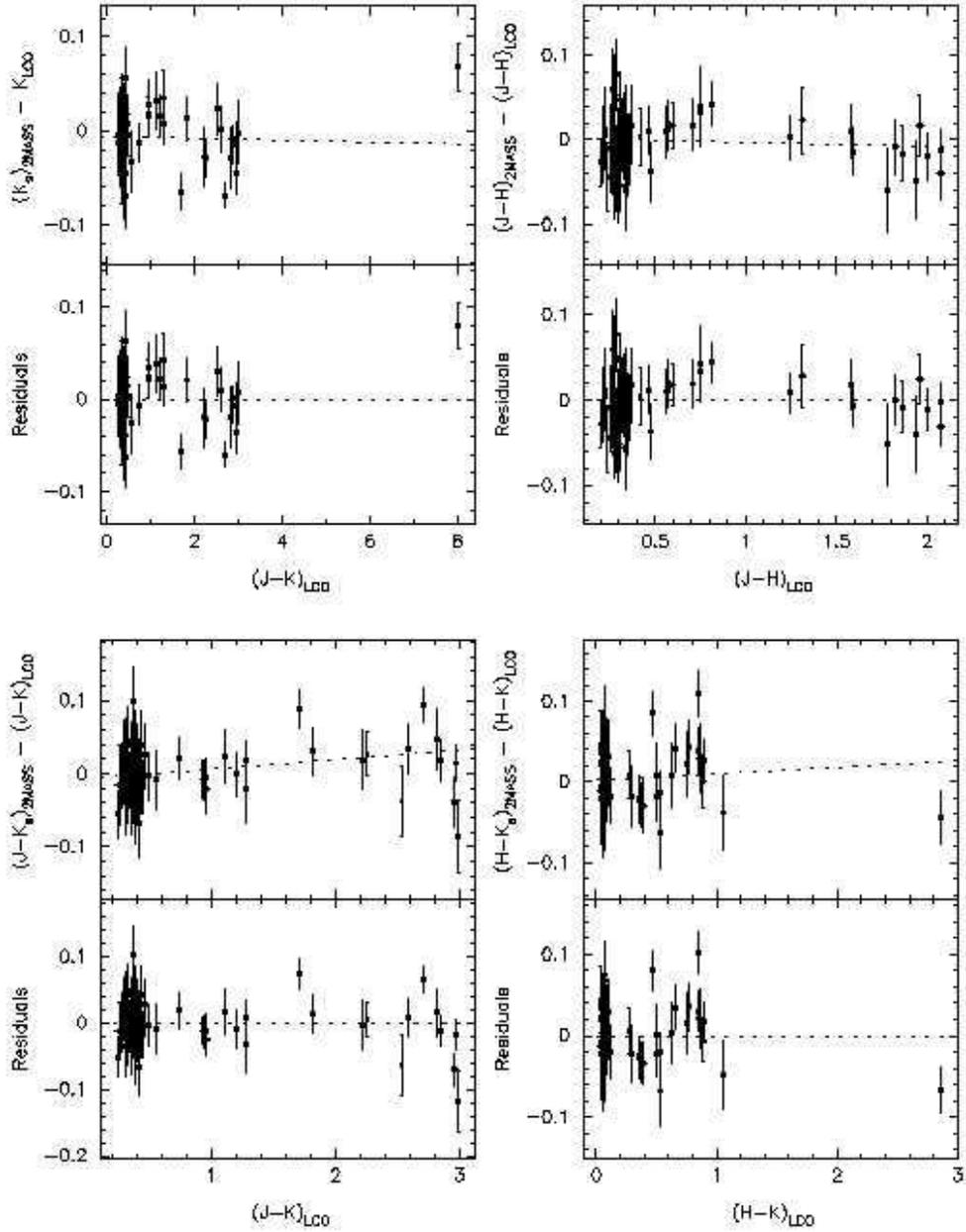}{8.8}{8.1}{0.0}{0.8}{0.8}{0}
\caption{
  Same as Figure~\ref{fig:aao}, except for the LCO photometric 
  system with the $K$-band photometry as summarized in \citet{Persson98}.
  Comparison to the LCO \KB-band photometry is provided in 
  Figure~\ref{fig:lco_s}.
  \label{fig:lco}
}
\end{figure}
\clearpage

\begin{figure}
\insertplot{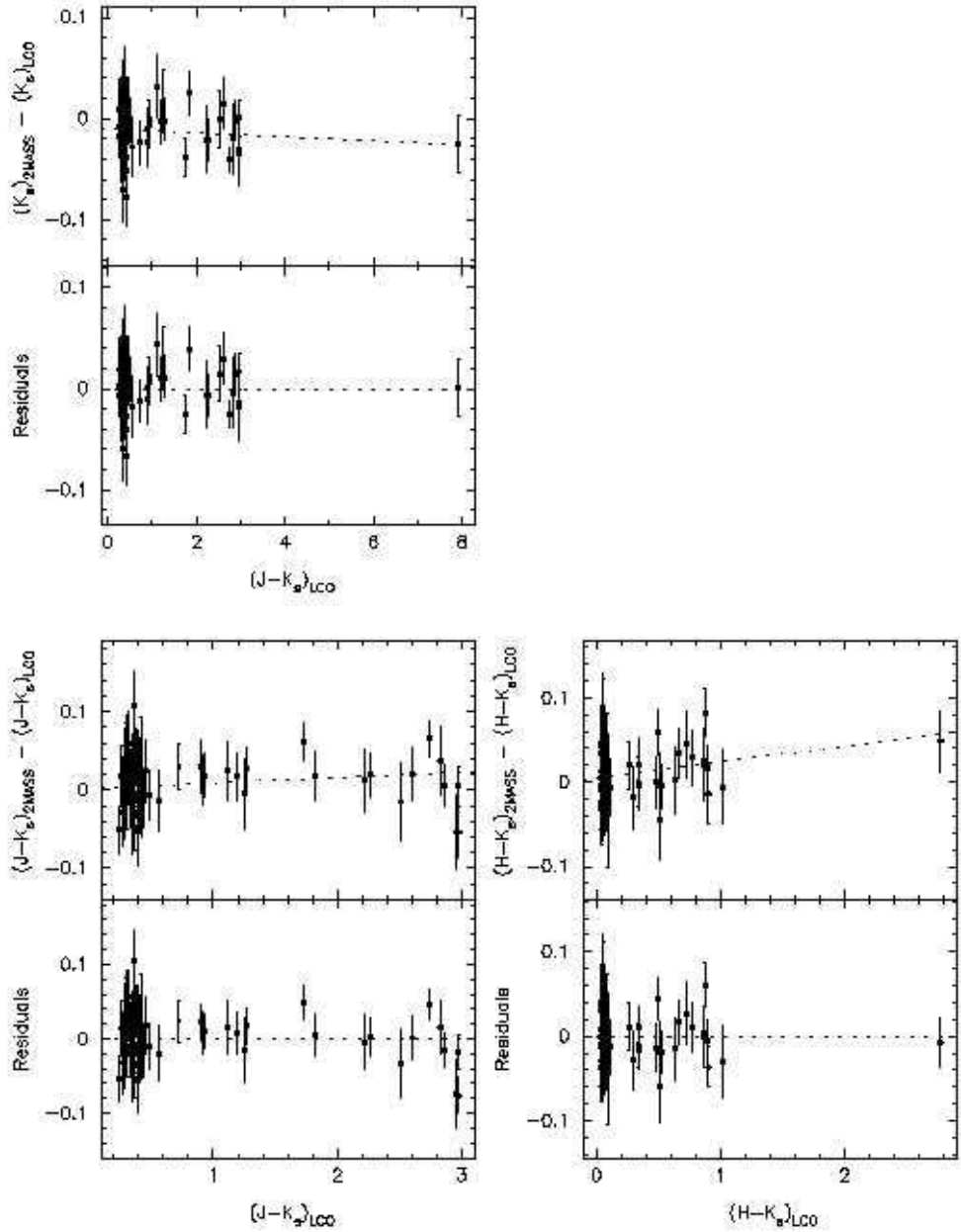}{8.8}{8.1}{0.0}{0.8}{0.8}{0}
\caption{
  Same as Figure~\ref{fig:aao}, except for the LCO (Persson) photometric 
  system with the \KB-band photometry as summarized in \citet{Persson98}.
  The $J-H$ data for the LCO system and comparison of the LCO $K$-band 
  photometry with 2MASS are shown in Figure~\ref{fig:lco}.
  \label{fig:lco_s}
}
\end{figure}
\clearpage

\begin{figure}
\insertplot{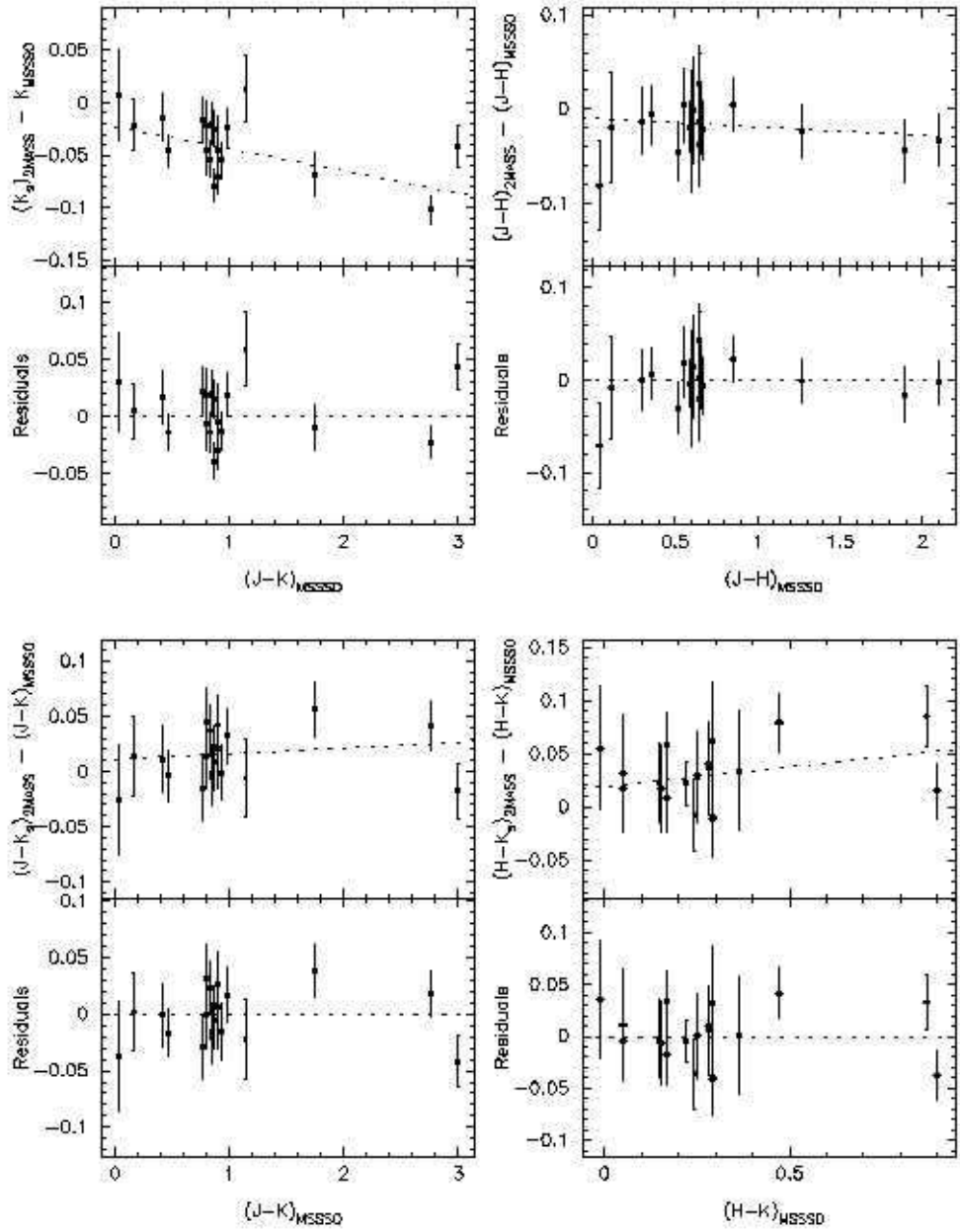}{8.8}{8.1}{0.0}{0.8}{0.8}{0}
\caption{
  Same as Figure~\ref{fig:aao}, except for the MSSSO photometric
  system. The MSSSO photometry is from \citet{M94}.
  \label{fig:mssso}
}
\end{figure}
\clearpage

\begin{figure}
\insertplot{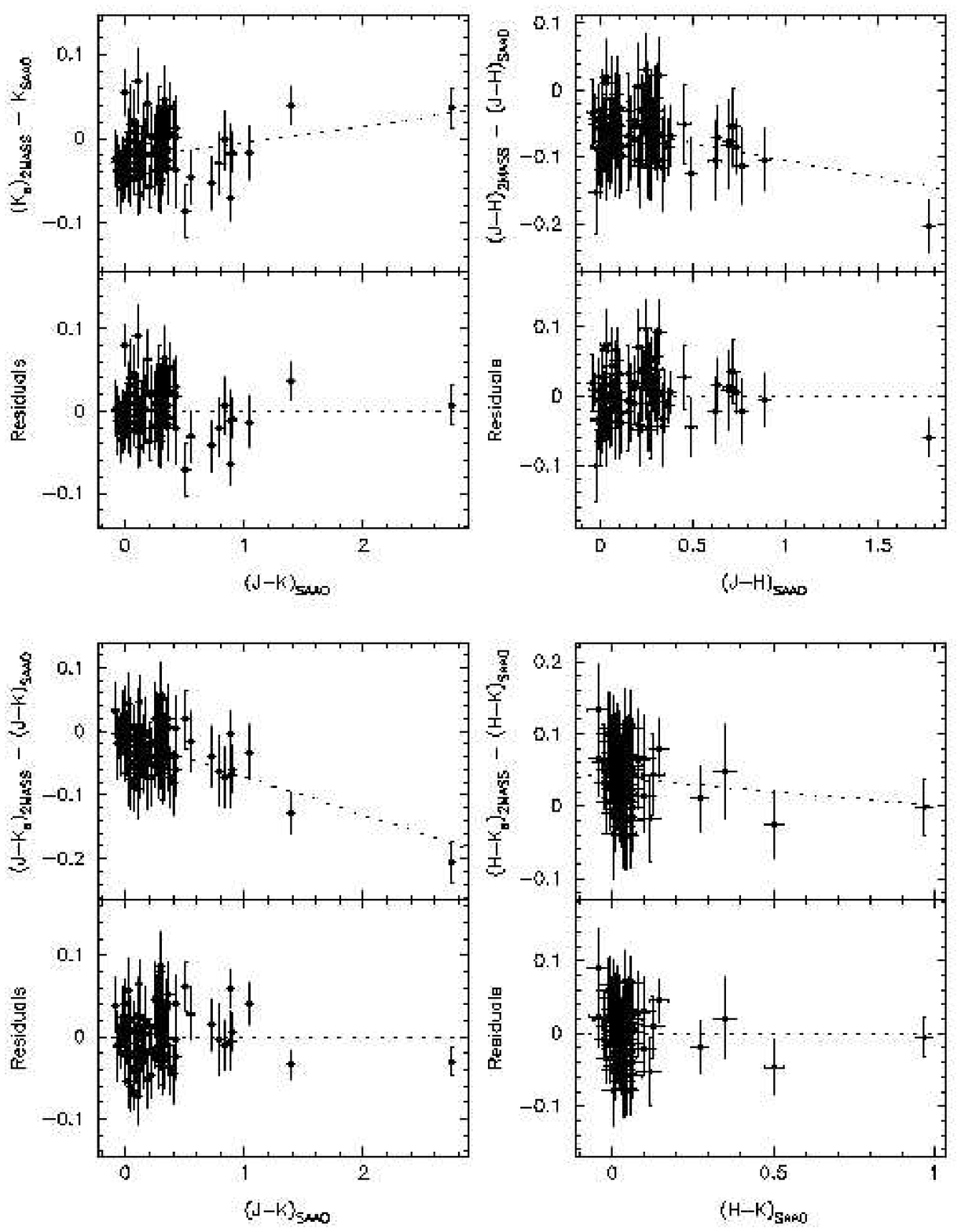}{8.8}{8.1}{0.0}{0.8}{0.8}{0}
\caption{
  Same as Figure~\ref{fig:aao}, except for the SAAO photometric
  system. The SAAO photometry is from \citet{Carter90} and \citet{Carter95}.
  \label{fig:saao}
}
\end{figure}
\clearpage

\begin{figure}
\insertplot{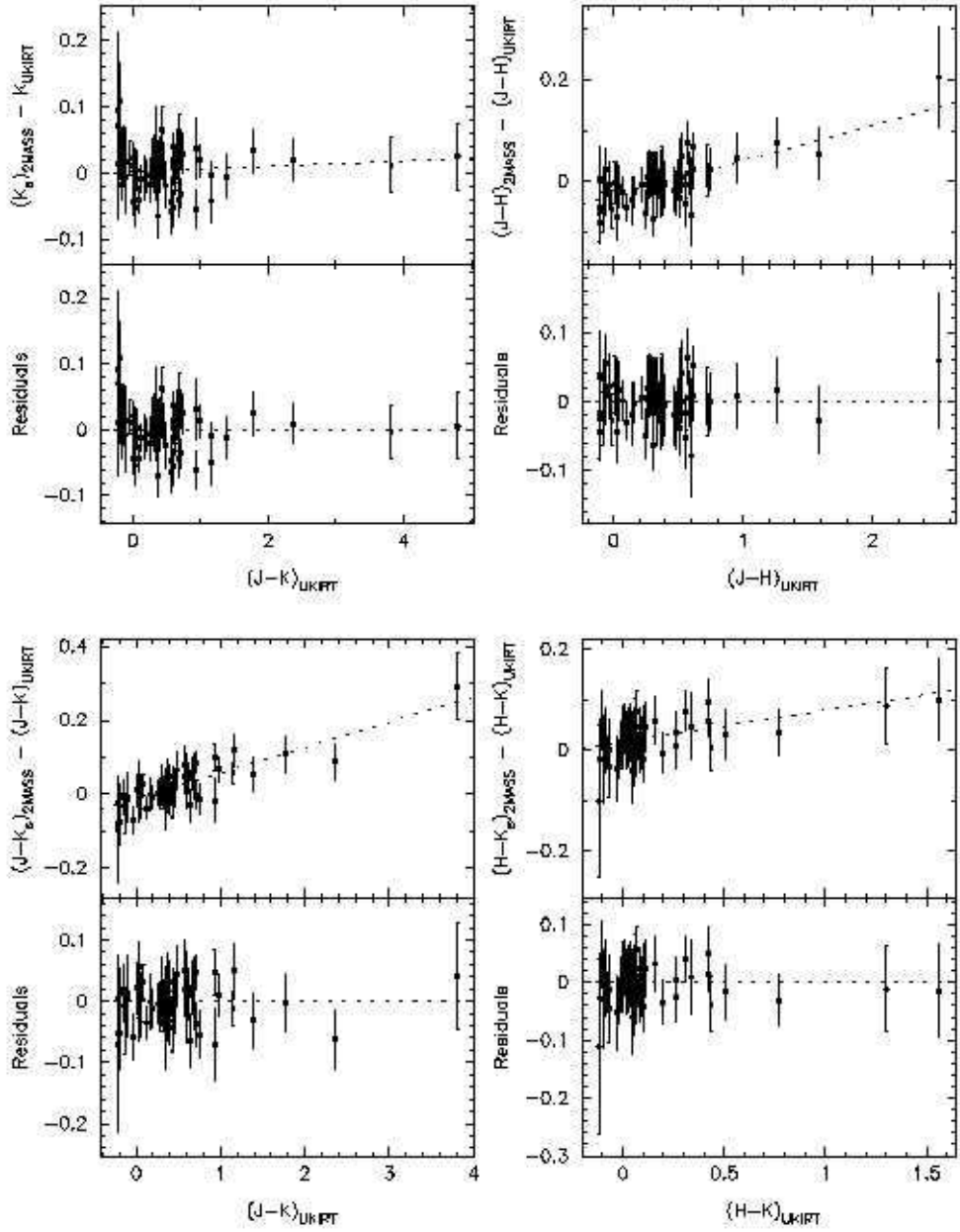}{8.8}{8.1}{0.0}{0.8}{0.8}{0}
\caption{
  Same as Figure~\ref{fig:aao}, except for UKIRT
  photometric system. The UKIRT standard star photometry is
  from \citet[see also Casali \& Hawarden~1992]{H00}.
  \label{fig:ukirt}
}
\end{figure}
\clearpage

\clearpage

%\begin{table}
%\dummytable\label{tbl:standards}
%\dummytable\label{tbl:q}
%\end{table}

\end{document}

%% file: table1.tex
%\documentstyle[/scr/jmc/tex/aastex_old/aj_pt4]{article}

%\begin{document}

%\newcommand{\etal}{et al.}

%\tablenum{1}
%\pagestyle{empty}

\begin{deluxetable}{llr}
\tablewidth{250pt}
\tablecaption{Photometric Systems\label{tbl:standards}}
\tablehead{
\colhead{System}  & 
\colhead{Reference}    & 
\colhead{Nstars}
}
\startdata
AAO     & Allen \& Cragg 1983      &  4\\
        & Elias \etal~1983         & 10\\
ARNICA  & Hunt  \etal~1988         & 65\\
CIT     & Elias \etal~1982         & 33\\
        & Elias \etal~1983         &  8\\
DENIS   & Epchtein \etal~1999      & 190\\
ESO     & van der Bliek \etal~1996 & 56\\
LCO     & Persson \etal~1998       & 82\\
MSSSO   & McGregor 1994            & 20\\
SAAO    & Carter 1990              & 29\\
        & Carter \& Meadows 1995   & 65\\
UKIRT   & Hawarden \etal~2000      & 72\\
\enddata
\end{deluxetable}

%\end{document}

%% file: table2.tex
%\documentstyle[/scr/jmc/tex/aastex_old/aj_pt4]{article}

%\begin{document}

%\newcommand{\etal}{et al.}

%\tablenum{1}
%\pagestyle{empty}

\begin{deluxetable}{lccccccccc}
\tablewidth{452pt}
\tablecaption{Goodness-of-Fit Parameters\label{tbl:q}}
\tablehead{
\colhead{System}                 & 
\multicolumn{4}{c}{$\chi^2_\nu$} & 
\colhead{}                       & 
\multicolumn{4}{c}{$q$} \\
\cline{2-5}
\cline{7-10}
\colhead{}      & 
\colhead{$K$}   &
\colhead{$J-H$} &
\colhead{$J-K$} &
\colhead{$H-K$} &
\colhead{}      &
\colhead{$K$}   &
\colhead{$J-H$} &
\colhead{$J-K$} &
\colhead{$H-K$}
}
\startdata
AAO         & 1.6 & 0.4 & 1.2 & 0.8 && 0.08  & 0.95 & 0.26 & 0.68\\
ARNICA      & 1.1 & 0.5 & 0.5 & 0.5 && 0.36  & 1.00 & 1.00 & 1.00\\
CIT         & 1.8 & 0.9 & 1.3 & 0.8 && 0.002 & 0.60 & 0.10 & 0.84\\
DENIS       & 2.4 & --- & 2.0 & --- && $5\times10^{-23}$ & --- & $5\times10^{-15}$ & ---\\
ESO         & 1.6 & 0.8 & 1.1 & 0.8 && 0.003 & 0.82 & 0.33 & 0.80\\
Koornneef\tablenotemark{a} & 0.5 & 0.2 & 0.2 & 0.1 && 0.99  & 1.00 & 1.00 & 1.00\\
LCO ($K$)   & 1.2 & 0.5 & 0.8 & 0.8 && 0.16  & 1.00 & 0.88 & 0.95\\
LCO ($K_s$) & 0.7 & --- & 0.6 & 0.5 && 0.99  &  --- & 1.00 & 1.00\\
MSSSO       & 1.6 & 0.4 & 0.7 & 0.7 && 0.05  & 0.99 & 0.79 & 0.84\\
SAAO        & 0.8 & 0.5 & 0.6 & 0.5 && 0.85  & 1.00 & 1.00 & 1.00\\
UKIRT       & 1.2 & 0.5 & 0.5 & 0.4 && 0.15  & 1.00 & 1.00 & 1.00\\
\enddata
\tablenotetext{a}{Goodness-of-fit-results are for the SAAO vs. Koornneef fit
                  described in Appendix B.}
\end{deluxetable}
%\end{document}